\documentstyle[12pt,aps,epsfig]{revtex}
\begin{document}

\title{Study of $^{44}$Ti in a Mixed--Symmetry Basis}

\author{V. G. Gueorguiev\footnote{Corresponding author. Email address:
vesselin@phys.lsu.edu} , J. P. Draayer,}
\address{\it
Department of Physics and Astronomy, Louisiana State University,\\ 
Baton Rouge, Louisiana 70803 \\ }

\author{W. E. Ormand}
\address{\it 
Physics Directorate, L-414, Lawrence Livermore National Laboratory,\\
P.O. Box 808, Livermore, CA 94551 \\ }

\author{C. W. Johnson}
\address{\it
Department of Physics, San Diego State University, \\ 
San Diego, CA 92182-1233}

\date{\today}
\maketitle

\begin{abstract}
The structure of $^{44}$Ti is studied in an oblique-basis that includes
spherical and SU(3) shell-model basis states. The results show that the
oblique-basis concept is applicable, even though the strong spin-orbit
interaction, which breaks the SU(3) symmetry, generates significant splitting
of the single-particle levels. Specifically, a model space that includes a
few SU(3) irreducible representations (irreps), namely, the leading (12,0)
and next to the leading (10,1) irreps -- including spin $S=0$ and $1$
configurations of the latter, plus spherical shell-model configurations
(SSMC) that have at least two valence nucleons confined to the $f_{7/2}$
orbit -- the SM(2) case, yield results that are comparable to SSMC with at
least one valence nucleon confined to the $f_{7/2}$ orbit -- the SM(3) case.
\end{abstract}



\textit{Introduction} --
In a previous paper we introduced the mixed-symmetry, oblique-basis concept
and demonstrated its applicability in the $sd$-shell using $^{24}$Mg as an
example \cite{VGG 24MgObliqueCalculations}. The successful description of
$^{24}$Mg can be understood in terms of the comparable importance of
single-particle excitations that are described most naturally in terms of
spherical shell-model configurations (SSMC), and collective quadrupole
excitations that are best described by the SU(3) shell model. An important
element to the success of the theory for lower $sd$-shell nuclei is the fact
that SU(3) is a reasonably good symmetry \cite{Elliott's SU(3) model}; that
is, for lower $sd$-shell nuclei SU(3) is the dominant symmetry while the
spherical shell-model (SSM) scheme plays an important but clearly more
recessive role.

In contrast with the $sd$-shell situation, for lower $pf$-shell nuclei
the strong spin-orbit splitting dominates the landscape and SU(3) symmetry
is badly broken \cite{VGG SU(3)andLSinPF-ShellNuclei}. Therefore, one might
anticipated that adding the leading and next to the leading SU(3) irreps to
the dominate SSM configurations for lower $pf$-shell nuclei might not add
much value to the analysis. Nevertheless, our calculations for $^{44}$ Ti
show that the addition of leading SU(3) irreps speeds the convergence; that
is, whereas in this case the spherical shell-model (SSM) is clearly dominant
and SU(3) is recessive, the latter remains important and signals that the
oblique-basis remains an important concept even in situations where one of
the symmetries is rather badly broken.

Here we consider oblique-basis calculations for $^{44}$Ti using the KB3
interaction \cite{KB3 interaction}. We confirm that the spherical shell
model (SSM) provides a significant part of the low-energy wave functions
within a relatively small number of SSMC while an SU(3) shell-model scheme
with only few SU(3) irreps is unsatisfactory. This is the opposite of the
situation in the $sd$-shell. Since the SSM yields relatively good results
with SM(2), which includes SSMC of up to two valence nucleons free to move in 
any of $pf$-shell orbitals, combining the two basis sets yields even better
results with only a very small increase in the overall size of the model space.
In particular, results in a SM(2)+SU(3) model space (47.7\% + 2.1\% of the full
$pf$-shell space) are comparable with SM(3) results (84\% of the full
$pf$-shell space). Therefore, as for the $sd$-shell, combining a few SU(3)
irreps with SM(2) configurations yields excellent results, such as correct
spectral structure, good ground-state energy, and an overall improved
structure of the wave functions.

\textit{Model Space} --
$^{44}$Ti consists of 2 valence protons and 2 valence neutrons in the $pf$%
-shell. The SU(3) basis includes the leading irrep (12,0) with $M_{J}=0$
dimensionality 7, and the next to the leading irrep (10,1). The (10,1)
occurs three times, once with $S=0$ (dimensionality 11) and twice with $S=1$
(dimensionality $2\times 33=66$). All three (10,1) irreps have a total
dimensionality of 77. The total $M_{J}=0$ dimensionality of (12,0) and
(10,1) is therefore 84. In Table \ref{TableTi44} we summarize the
dimensionalities involved in our calculation.

\begin{figure}[tbp]
\begin{center}
\leavevmode
\epsfxsize = 14.5cm 
\epsffile{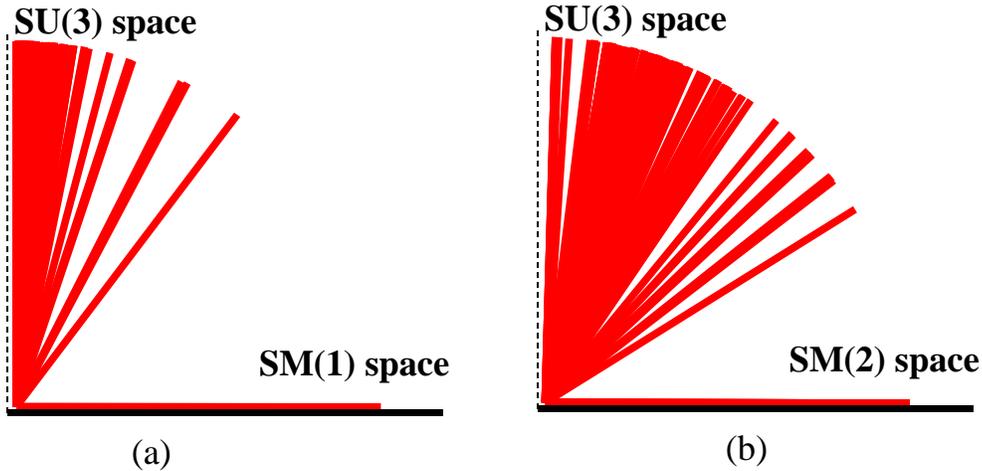}
\end{center}
\caption{Orthogonality of the basis vectors in the oblique geometry. The
SU(3) space consist of (12,0)\&(10,1) basis vectors. The shell-model spaces
(SM($n$) with $n=$1 and 2) is indicated by a horizontal line. (a) SM(1) and
the leading SU(3) basis vectors; there are two SU(3) vectors that lie in the
SM(1) space. (b) SM(2) and the leading SU(3) basis vectors; there are twelve
SU(3) vectors that lie in the SM(2) space.}
\label{SU3SM1SM2}
\end{figure}

Within oblique bases type calculations one expects some linearly
dependent vectors \cite{VGG 24MgObliqueCalculations}. In our example, there
is one redundant vector in the SM(2)+(12,0) space, two in SM(3)+(12,0) and
SM(1)+(12,0)\&(10,1) spaces, twelve in SM(2)+(12,0)\&(10,1) space, and
thirty-three in the SM(3)+(12,0)\&(10,1) space. Each linearly dependent
vector is handled as discussed previously \cite{VGG 24MgObliqueCalculations}.
The structure of the oblique-basis model space is shown in Fig.
\ref{SU3SM1SM2}.

\begin{table}[tbp]
\begin{tabular}{lrrrrrrr}
Model space & (12,0) & \&(10,1) & SM(0) & SM(1) & SM(2) & SM(3) & FULL \\
\hline
dimension & 7 & 84 & 72 & 580 & 1908 & 3360 & 4000 \\
dimension \% & 0.18 & 2.1 & 1.8 & 14.5 & 47.7 & 84 & 100 \\
\end{tabular}
\caption{Labels and M$_{J}$=0 dimensions of various model spaces for
$^{44}$Ti. The leading SU(3) irrep is (12,0); \&(10,1) implies that the
three (10,1) irreps (one with $S=0$ and M$_{J}$=0 dimensionality 11 and
two with $S=1$ and M$_{J}$=0 dimensionality 33 each) are included along
with the leading irrep (12,0). SM(n) is a spherical shell-model basis with
n valence particles anywhere within the full $pf$-shell when the remaining
valence particles are being confined to the $f_{7/2}$ orbit.}
\label{TableTi44}
\end{table}

\begin{figure}[tbh]
\begin{center}
\leavevmode
\epsfxsize = 14.5cm 
\epsffile{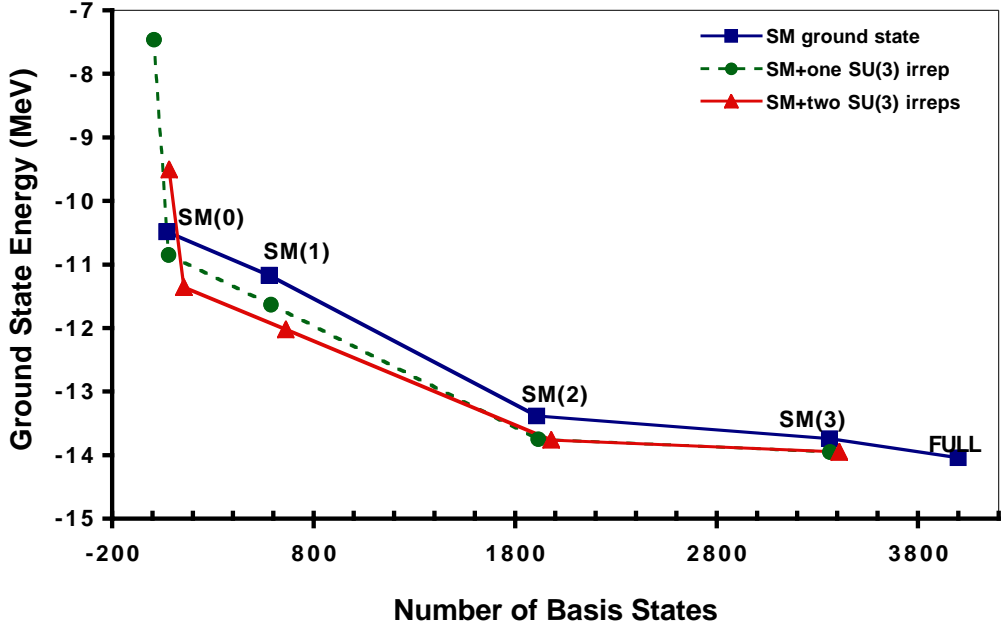}
\end{center}
\caption{Ground-state energy for $^{44}$Ti as a function of the various
model spaces. The SU(3) irreps used in the calculations are (12,0) and
(10,1).}
\label{Ti44DimConv}
\end{figure}

\textit{Ground-State Energy} --
The convergence of the ground-state energy as function of the model space
dimension is shown in Fig. \ref{Ti44DimConv}. The oblique-basis calculation
of the ground-state energy for $^{44}$Ti does not appear to be as
impressive as for $^{24}$Mg \cite{VGG 24MgObliqueCalculations}, especially
when we compare SM(2) with SM(1) plus SU(3). The calculated ground-state
energy for the SM(1)+(12,0)\&(10,1) space is $0.85$ MeV below the calculated
energy for the SM(1) space. In contrast, the ground-state energy for SM(2) is
$2.2$ MeV below the SM(1) result. Adding the two SU(3) irreps to the
SM(1) basis increases the size of the space from 14.5\% to 16.6\% of the full
space. This is a 2.1\% increase, while going from the SM(1) to SM(2) involves
an increase of 33.2\%. However, adding the SU(3) irreps to the SM(2) basis
gives ground-state energy of $-13.76$ MeV which is compatible to the SM(3)
result of $-13.74$ MeV. Therefore, adding the SU(3) to the SM(2) increases
the model space from 47.7\% to 49.8\% and gives results that are slightly
better than the results for SM(3) model space that is 84\% of the full space.

\begin{figure}[tbh]
\begin{center}
\leavevmode
\epsfxsize = 14.5cm 
\epsffile{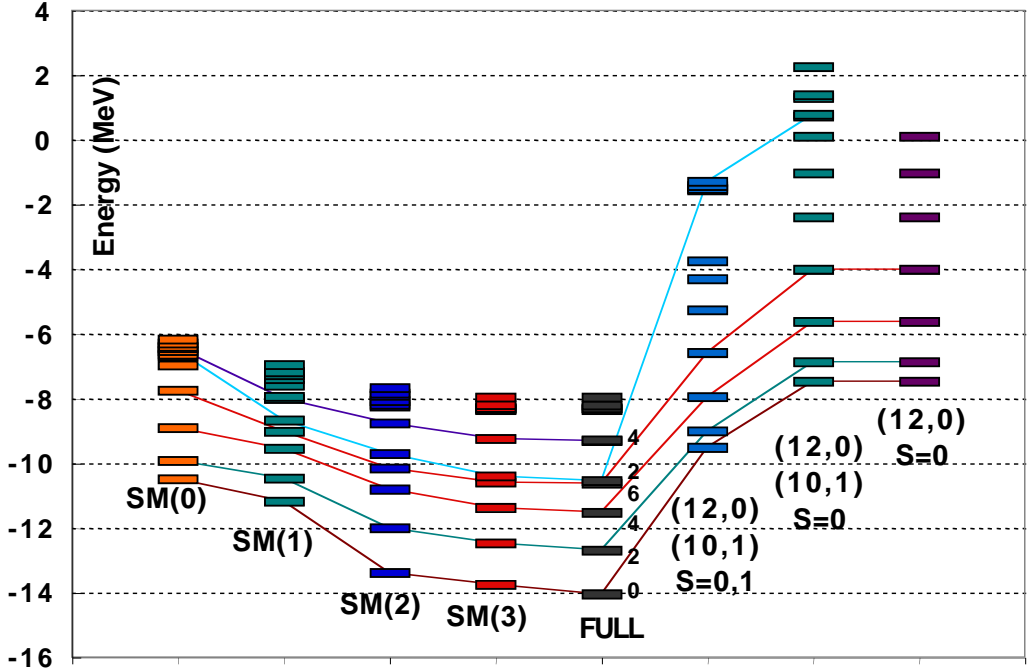}
\end{center}
\caption{Structure of the energy levels for $^{44}$Ti for different
calculations. Pure $m$-scheme spherical-basis calculations are on the
left-hand side while pure SU(3)-basis calculations are on the right-hand
side. The spectrum from the FULL space calculation is in the center.}
\label{Ti44LevelStructure}
\end{figure}

\textit{Low-Lying Energy Spectrum} --
For $^{24}$Mg the position of the K=2 band head is correct for SU(3)-type
calculations but not for the low-dimensional SM(n) calculations \cite{VGG
24MgObliqueCalculations}. For $^{44}$Ti we find the opposite scenario which
is shown in Fig. \ref{Ti44LevelStructure}. In this case the SM(n)-type
calculations reproduce the position of the K=2 band head while SU(3)-type
calculations put the K=2 band head too high. Furthermore, the low-energy
levels for the SU(3) case are higher than the SM(n) case, which is a
scenario that may not produce the necessary mixing of the levels that would
lead to a better spectral structure. It is important to note that basis states
with spin other than zero are essential to achieve a proper description of the
low energy spectrum.  This can be seen very clearly for $^{44}$Ti in Fig.
\ref{Ti44LevelStructure} where the addition of the $S=1$ (10,1) irreps to the
(12,0) and (10,1) $S=0$ pair increases the binding energy by 2 MeV.

\begin{figure}[tbh]
\begin{center}
\leavevmode
\epsfxsize = 14.5cm 
\epsffile{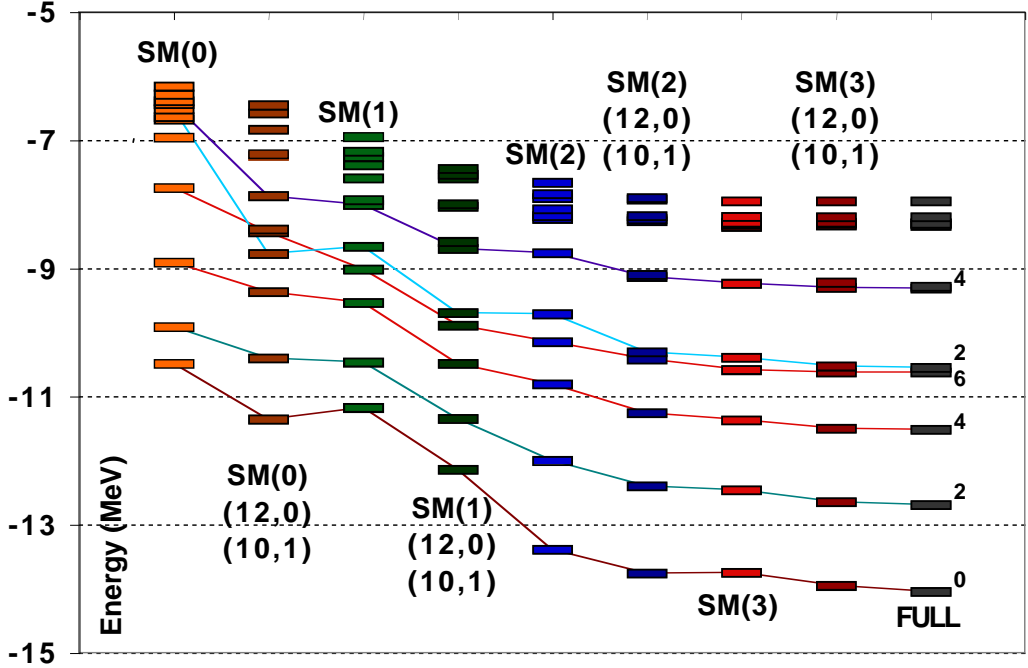}
\end{center}
\caption{Structure of energy levels for $^{44}$Ti for different oblique-basis
calculations using leading and next to the leading irreps with spin $S=0$ and
$1$. Pure $m$-scheme spherical-basis calculations are included for
comparison.}
\label{SMSU3Spectrum}
\end{figure}

Overall, the spectral structure in the oblique-basis calculation is good
with the SM(2)+(12,0)\&(10,1) spectrum ($\approx $50\% of the full space)
delivering results that are comparable to the SM(3) case (84\%). Additional
details are shown in Fig. \ref{SMSU3Spectrum}.

\begin{figure}[tbh]
\begin{center}
\leavevmode
\epsfxsize = 14.5cm 
\epsffile{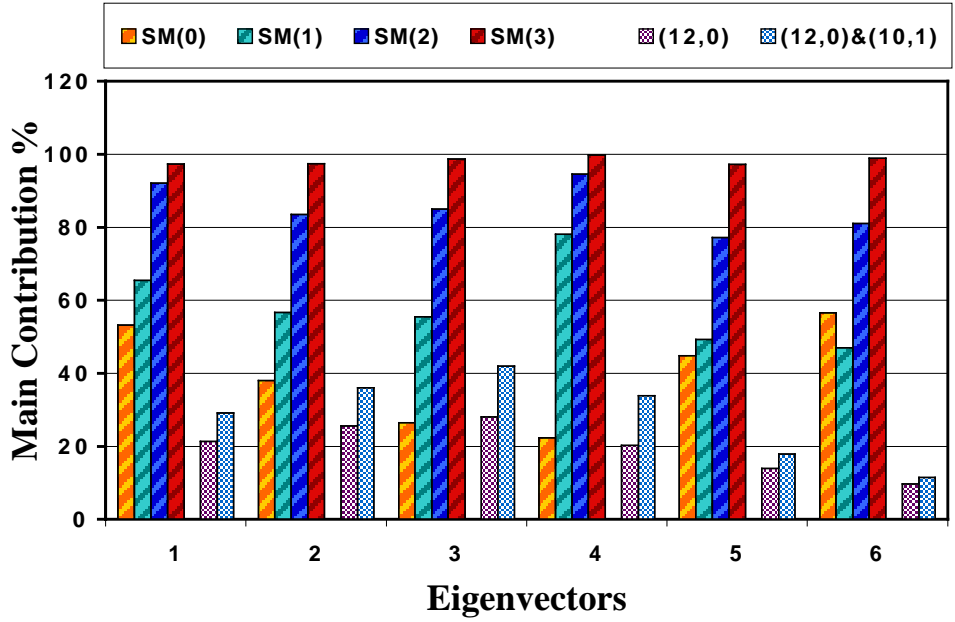}
\end{center}
\caption{Wave function overlaps of the FULL $pf$-shell states of $^{44}$Ti
with pure spherical shell model and SU(3) type calculations. The first four
bars represent the SM(0), SM(1), SM(2), and SM(3) calculations, the next two
bars represent the SU(3) calculations.}
\label{Ti44Overlaps}
\end{figure}

\textit{Overlaps with Exact States} --
The overlap of the SU(3)-type eigenstates with the exact (full $pf$-shell
model space) results are not as large as in the $sd$-shell, often running
less than 40\%. This result was also reported in a comparative study of SU(3)
and projected Hartree-Fock \cite{PHF and SU(3)} calculations.  The SM(n)
results are considerably better with SM(2)-type calculations yielding on
average more than an 85\% overlap with the exact states while the results for
SM(3) show overlaps greater than 97\%, which is consistent with the fact that
SM(3) covers 84\% of the full space. This situation is shown in Fig.
\ref{Ti44Overlaps}.

\begin{figure}[tbh]
\begin{center}
\leavevmode
\epsfxsize = 14.5cm 
\epsffile{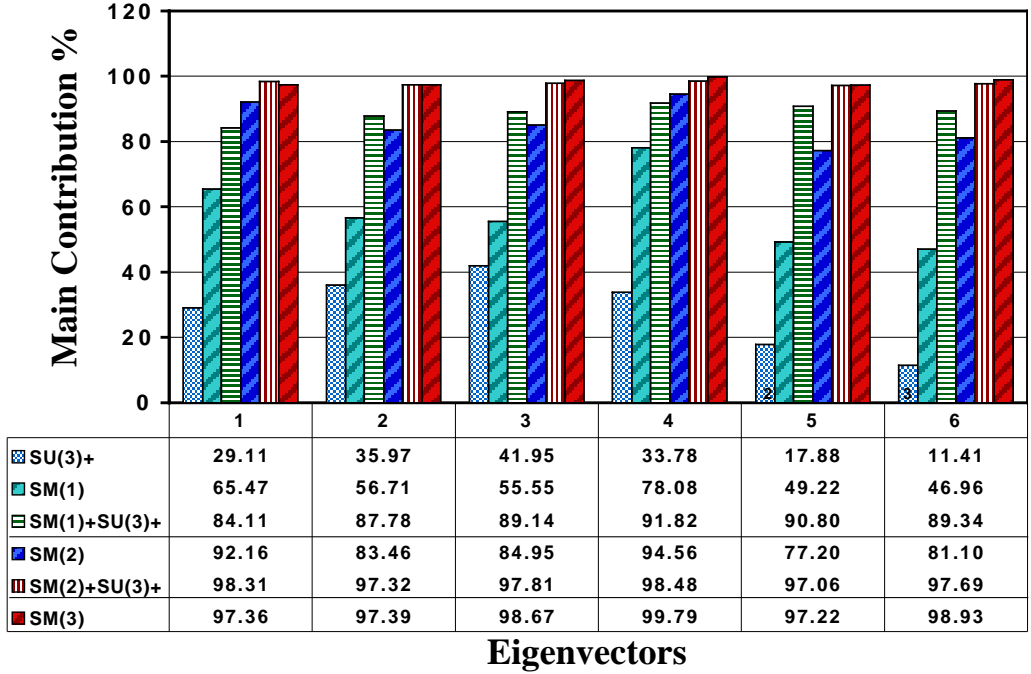}
\end{center}
\caption{Selected overlaps of SM(n), SU(3), and oblique-basis results with the
exact full $pf$-shell eigenstates for $^{44}$Ti. Here SU(3)+ denotes the
(12,0)\&(10,1) SU(3) irreps.}
\label{SelectedOverlaps}
\end{figure}

On the other hand,  as shown in Fig. \ref{SelectedOverlaps}, SM(2) plus
(12,0)\&(10,1)-type calculations yield results in about 50\% of the
full-space that are as good as those for SM(3) which is 84\% of the full
$pf$-shell model space.  Notice that the SM(1)+(12,0)\&(10,1) overlaps are
often bigger than the SM(2) overlaps as shown in Fig. \ref{SelectedOverlaps}.

\textit{Conclusion} --
For $^{44}$Ti, combining a few SU(3) irreps with SM(2) configurations
increases the model space only by a small ($\approx $2.3\%) amount but
results in better overall results: a lower ground-state energy, the correct
spectral structure (particularly the position of the K=$2^{+}$ band head), and
wave functions with a larger overlap with the exact results. The
oblique-bases SM(2)+(12,0)\&(10,1) for $^{44}$Ti ($\approx $50\% of the full
space) yields results that are comparable with SM(3) results ($\approx $84\%
of the full space). In short, the oblique-basis scheme works well for
$^{44}$Ti but in contrast with $^{24}$Mg, the SSM configurations are dominant
when SU(3) is  recessive.

\begin{acknowledgements}
We acknowledge support provided by the U.S. Department of Energy under Grant
No. DE-FG02-96ER40985, and the U.S. National Science Foundation under Grant
Nos. PHY-9970769 and PHY-0140300.
\end{acknowledgements}

\end{document}